\documentclass{article}

\PassOptionsToPackage{numbers, compress}{natbib}

\usepackage[preprint]{neurips_2022}

\usepackage[utf8]{inputenc} 
\usepackage[T1]{fontenc}    
\usepackage{hyperref}       
\usepackage{url}            
\usepackage{booktabs}       
\usepackage{amsfonts}       
\usepackage{nicefrac}       
\usepackage{microtype}      
\usepackage{xcolor}         
\usepackage{graphicx}
\usepackage{subfigure}
\usepackage{tipa}

\title{Dataset Distillation for Medical Dataset Sharing}

\author{%
  Guang Li \\
  Hokkaido University \\
  Sapporo, Japan \\
  \texttt{guang@lmd.ist.hokudai.ac.jp} \\
  \And
  Ren Togo \\
  Hokkaido University \\
  Sapporo, Japan \\
  \texttt{togo@lmd.ist.hokudai.ac.jp} \\
  \AND
  Takahiro Ogawa \\
  Hokkaido University \\
  Sapporo, Japan \\
  \texttt{ogawa@lmd.ist.hokudai.ac.jp} \\
  \And
  Miki Haseyama \\
  Hokkaido University \\
  Sapporo, Japan \\
  \texttt{mhaseyama@lmd.ist.hokudai.ac.jp} \\
}

\begin{document}

\maketitle

\begin{abstract}
Sharing medical datasets between hospitals is challenging because of the privacy-protection problem and the massive cost of transmitting and storing many high-resolution medical images.
However, dataset distillation can synthesize a small dataset such that models trained on it achieve comparable performance with the original large dataset, which shows potential for solving the existing medical sharing problems.
Hence, this paper proposes a novel dataset distillation-based method for medical dataset sharing.
Experimental results on a COVID-19 chest X-ray image dataset show that our method can achieve high detection performance even using scarce anonymized chest X-ray images.
\end{abstract}
\section{Introduction}
The sharing of medical datasets is essential in enabling the cross-hospital flow of medical information and improving the quality of medical services~\cite{kumar2021integration}.
However, sharing healthcare datasets between different hospitals faces several thorny issues.
Firstly, privacy protection has been a severe issue hindering the process when sharing medical image datasets from different hospitals~\cite{kaissis2020secure}.
Second, sharing large-scale high-resolution medical image datasets increases transmission and storage costs~\cite{li2022compressed}.
Therefore, the solution to these problems will significantly promote the development of medical dataset sharing.
\par
Dataset distillation can synthesize a small dataset such that models trained on it achieve comparable performance with the original large dataset~\cite{wang2018datasetdistillation}.\footnote{https://github.com/Guang000/Awesome-Dataset-Distillation}
Although dataset distillation has been proposed for distilling some simple datasets, such as MNIST and CIFAR10, its effectiveness in high-resolution complex medical datasets has not yet been proved~\cite{zhao2021datasetcondensation}.
Medical dataset distillation may have potential advantages for solving the existing medical dataset sharing problems~\cite{li2020soft}.
For example, the size of distilled medical image datasets can be significantly compressed, and distilled images generated from noise are automatically anonymized~\cite{dong2022privacy}.
Therefore, it is desirable to explore the potential of dataset distillation for medical dataset sharing and contribute to real-world applications.
\par
\begin{figure}[t]
        \centering
        \includegraphics[width=13.5cm]{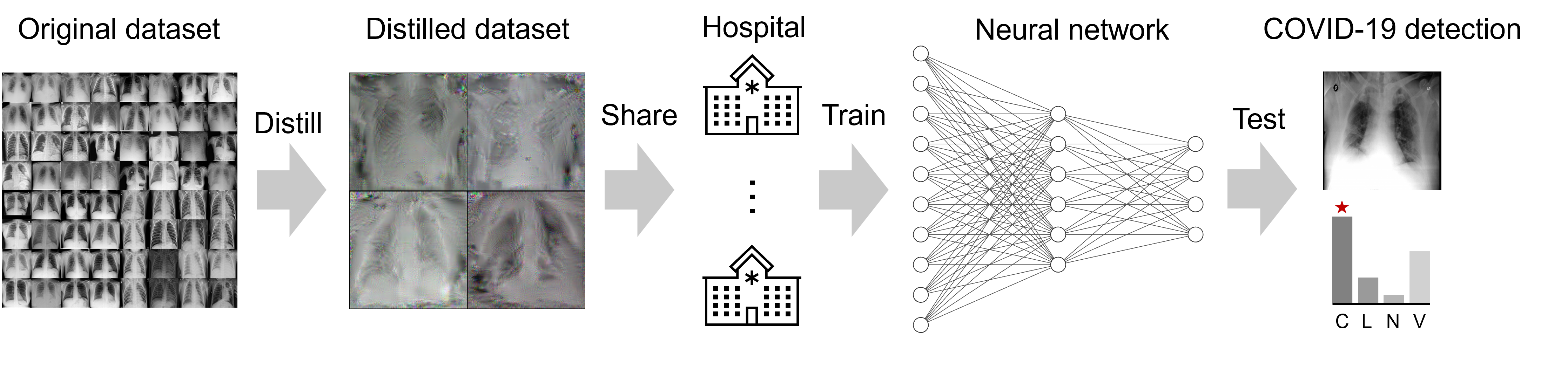}
        \caption{The concept of this study. Our method can improve the efficiency and security of the sharing of medical datasets between different hospitals.}
        \label{fig1}
\end{figure}

In this paper, we propose a novel dataset distillation-based method for medical dataset sharing.
COVID-19 and its variants have rapidly spread worldwide, influencing the health and life of billions of people~\cite{mofijur2021impact}.
X-ray is widely used in clinical because of its high speed and low cost.
Detecting COVID-19 from chest X-ray images is perhaps one of the fastest and easiest ways~\cite{minaee2020deep}.
However, sharing COVID-19 datasets between different hospitals also has the above-mentioned problems.
We perform experiments on a COVID-19 chest X-ray image dataset to prove the effectiveness of the proposed method.
Experimental results show that we can achieve high COVID-19 detection performance even using scarce anonymized chest X-ray images, hopeful of solving existing problems in medical dataset sharing.
The concept of this study is shown in Figure~\ref{fig1}.
\section{Methodology}
\begin{figure}[t]
        \centering
        \includegraphics[width=13.5cm]{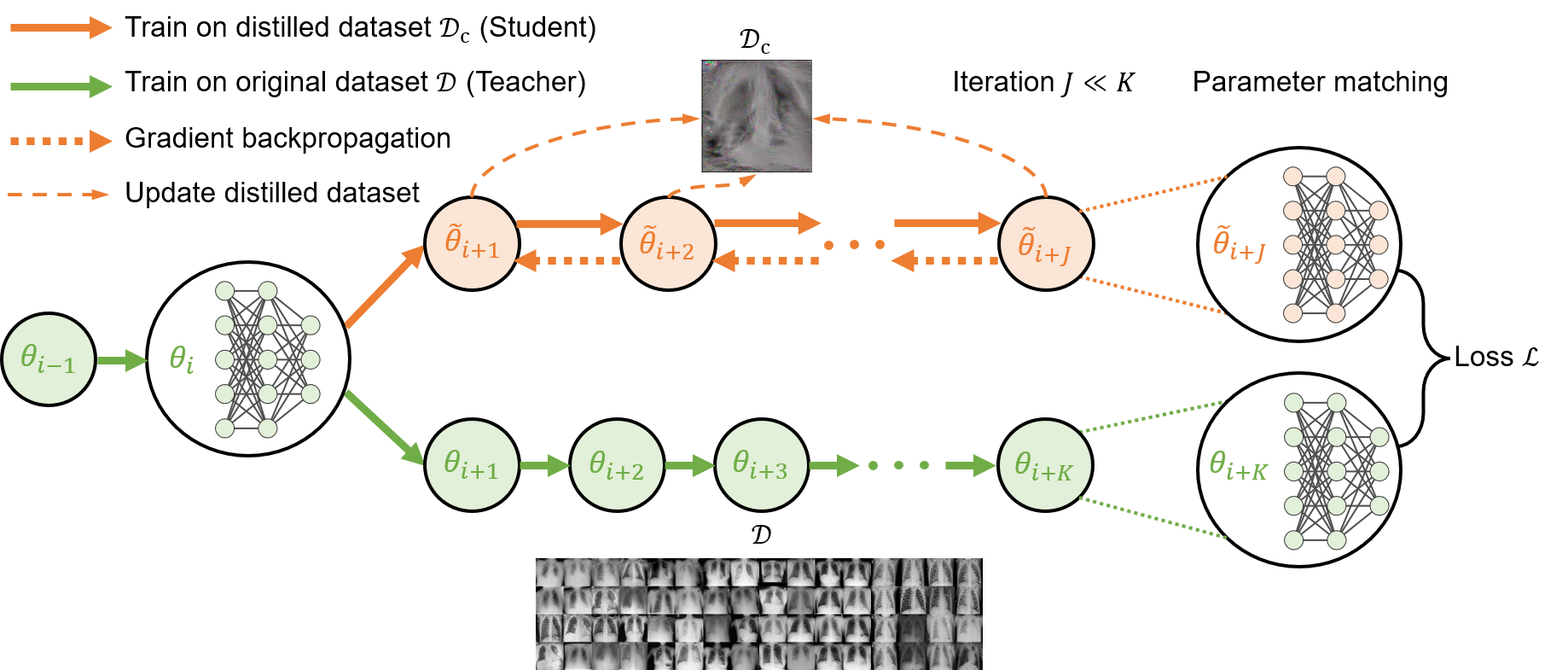}
        \caption{Overview of the proposed method.}
        \label{fig2}
\end{figure}

An overview of the proposed method is shown in Figure~\ref{fig2}.
The objective of our method is to have the parameters of the student network trained on the distilled dataset match the parameters of the teacher networks trained on the original dataset.
Before the distillation process, we first train $T$ teacher networks on the original COVID-19 dataset $\mathcal{D}$ and obtain their parameters~\cite{cazenavette2022dataset}.
These time sequences of parameters $\{\theta_{i}\}^{I}_{0}$ are defined as teacher parameters.
Also, network parameters trained on the distilled dataset $\mathcal{D}_{c}$ at each training step $i$ are defined as student parameters $\tilde{\theta}_{i}$.
Our method aims to distill chest X-ray images that induce network parameters similar to those learned from the original COVID-19 dataset (given the same initial values).
In the distillation process, student parameters are initialized as $\tilde{\theta}_{i}=\theta_{i}$ by sampled from
one of the teacher parameters at a random step $i$.
Then we perform gradient descent updates on the student parameters $\tilde{\theta}$ with respect to the cross-entropy loss $\ell$ of the distilled dataset $\mathcal{D}_{c}$ as follows:
\begin{equation}
\tilde{\theta}_{i+j+1} = \tilde{\theta}_{i+j} - \alpha\nabla\ell(\mathcal{A}(\mathcal{D}_{c});\tilde{\theta}_{i+j}),
\end{equation}
where $j$ and $\alpha$ represent the number of gradient descent updates and the trainable learning rate, respectively.
$\mathcal{A}$ represents a differentiable data augmentation module that can improve the distillation performance, which was proposed in~\cite{zhao2021differentiatble}.
Since the data augmentation used during distillation is differentiable, it can be propagated back through the augmentation layers to the distilled dataset.
Then we get the teacher parameters $\theta_{i+K}$ from $K$ gradient descent updates after the parameters used to initialize the student network.
The final loss $\mathcal{L}$ calculate the normalized $L_{2}$ loss between updated student parameters $\tilde{\theta}_{i+J}$ and teacher parameters $\theta_{i+K}$ as follows:
\begin{equation}
\mathcal{L} = \frac{|| \tilde{\theta}_{i+J}-\theta_{i+K} ||^{2}_{2}} {|| \theta_{i}-\theta_{i+K} ||^{2}_{2}},
\end{equation}
Finally, we minimize the loss $\mathcal{L}$ and backpropagate the gradient through all $J$ updates to the student network for obtaining the optimized distilled dataset $\mathcal{D}^{\ast}_{c}$.
Since the distilled chest X-ray images are generated from noise and have different distribution or visual similarities from the original images, they are automatically anonymized.
After obtaining the distilled dataset $\mathcal{D}^{\ast}_{c}$, we can share it with different hospitals and train neural networks for high-accuracy COVID-19 detection.
\section{Experiments}
The dataset used in our study has four classes, i.e., COVID-19, Lung Opacity, Normal, and Viral Pneumonia~\cite{rahman2021exploring}.
The number of images in each class is 3616, 6012, 10192, and 1345, respectively.
The resolution of chest X-ray images is 224 $\times$ 224, and we resized it to 112 $\times$ 112 for distillation or training networks.
The number of pre-trained teacher networks $T$ was set to 100. 
And we set the number of distilled images as 1, 2, 3, 5, 10, and 20 images per class.
The network structure used in this study is a simple ConvNet with depth-5 and width-128, which is often used in the dataset distillation task.
For comparative methods, we used several SOTA self-supervised learning methods, including SKD~\cite{li2022self}, BYOL~\cite{grill2020bootstrap}, SimSiam~\cite{chen2021exploring} and MAE~\cite{he2022masked}.
We also used transfer learning from ImageNet~\cite{deng2009imagenet} and training from scratch as baseline methods.
We randomly selected 42 images per class (1\% of the training set) for these comparative methods.
Except for the MAE method used ViT-Large~\cite{dosovitskiy2020vit}, all other methods used ResNet-50~\cite{he2022masked} as the backbone network.
\par
\begin{table}[t]
    \centering
    \caption{COVID-19 detection accuracy when using different numbers of distilled images. IPC denotes images per class.}
    \label{tab1}
    \begin{tabular}{l|cccccc|c}
    \hline
    IPC & 1 & 2 & 3 & 5 & 10 & 20 & Full Dataset \\\hline
    Accuracy
    & 52.5\% & 76.4\% & 77.0\% & 79.3\% & 82.2\% & \bfseries{82.7\%} & 88.9\% \\\hline
    \end{tabular}
\end{table}
\begin{table}[t]
    \centering
    \caption{COVID-19 detection accuracy of different methods.}
    \label{tab2}
    \begin{tabular}{l|ccccccc}
    \hline
    Method & \bfseries{Ours} & SKD & BYOL & SimSiam & MAE & Transfer & From Scratch \\\hline
    Accuracy
    & \bfseries{82.7\%} & 74.2\% & 68.3\% & 66.8\% & 62.3\% & 53.9\% & 28.4\% \\\hline
    \end{tabular}
\end{table}
\begin{figure}[t]
        \centering
        \includegraphics[width=10cm]{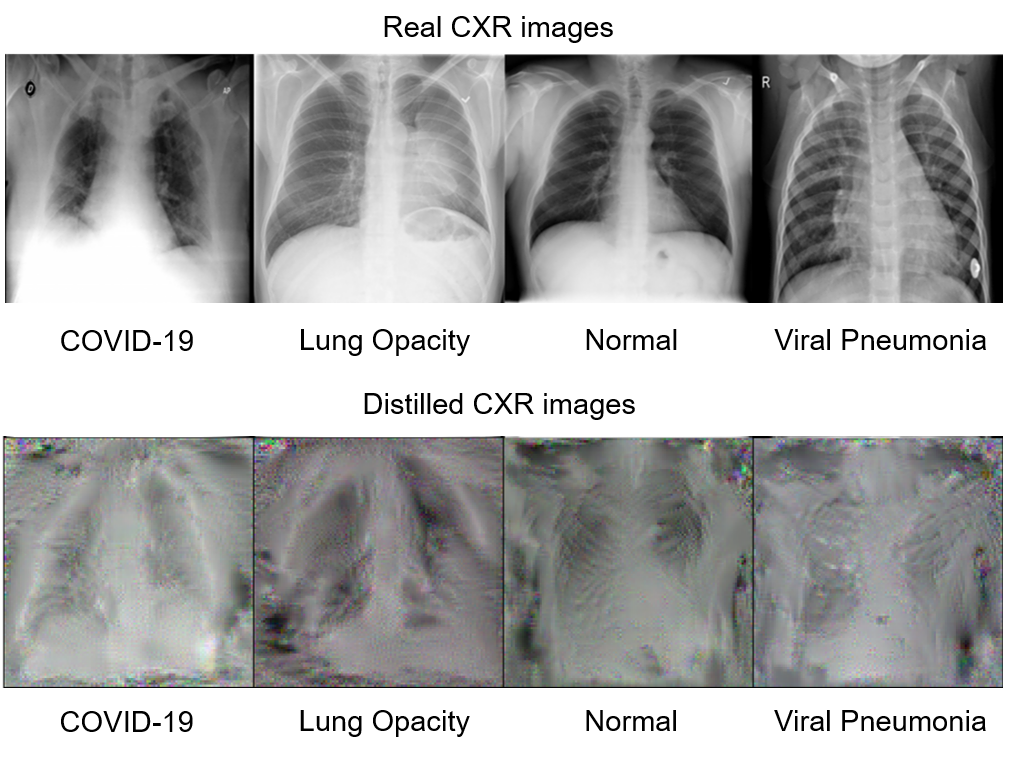}
        \caption{Examples of real and distilled images.}
        \label{fig3}
\end{figure}
The test accuracy of COVID-19 detection are shown in Tables~\ref{tab1} and~\ref{tab2}.
From Table~\ref{tab1}, we can see that the accuracy of our method increased accordingly as the number of distilled images grew.
We also show the upper bound accuracy of 88.9\% when training on the full dataset.
Even with a compression rate of 0.0047, no significant accuracy degradation is exhibited.
Table~\ref{tab2} shows that our method achieved high COVID-19 detection accuracy even when using scarce distilled chest X-ray images.
Furthermore, our method drastically outperformed other SOTA methods with a simpler network and fewer training images.
Figure~\ref{fig3} shows some examples of real and distilled images.
We can see that the distilled images are entirely visually different from the original images, which shows the anonymization effectiveness of the proposed method.
\section{Conclusion}
We have proposed a novel dataset distillation-based method for medical dataset sharing.
Since the size of the distilled medical image dataset has been significantly compressed and the images are also anonymized, the sharing of medical datasets between different hospitals will be more efficient and secure.
Experimental results on a COVID-19 chest X-ray image dataset show the advantage of our method compared to other SOTA methods.
\section*{Potential Negative Societal Impact}
The findings of this paper show the effectiveness of dataset distillation for medical dataset sharing.
Although the experimental results are promising, the proposed method should be verified on other medical datasets of different diseases for any potential bias.
Since the computational overhead of training and storing teacher parameters is relatively high, which may not necessarily be available in low-resource settings.
We also did some experiments to verify the usefulness of early distillation algorithms for medical images, but the results were not very effective and computationally intensive. 
Hence we did not present these results. 
Furthermore, verifying the validity of distilled medical images on other network structures and in terms of differential privacy will be our future work.
\section*{Acknowledgement}
This study was supported in part by AMED Grant Number JP21zf0127004, the Hokkaido University-Hitachi Collaborative Education and Research Support Program, and the MEXT Doctoral Program for Data-Related InnoVation Expert Hokkaido University (D-DRIVE-HU) Program. 
This study was conducted on the Data Science Computing System of Education and Research Center for Mathematical and Data Science, Hokkaido University.
\bibliographystyle{plainnat}
\bibliography{ref}
\end{document}